# Silicon Quantum Dots with Counted Antimony Donor Implants

M. Singh[*], J. L. Pacheco, D. Perry, E. Garratt, G. Ten Eyck, N. C. Bishop, J. R. Wendt, R. P. Manginell, J. Dominguez, T. Pluym, D. R. Luhman, E. Bielejec, M. P. Lilly and M. S. Carroll

[1]Sandia National Laboratories, Albuquerque, New Mexico 87185, USA

[2]Center for Integrated Nanotechnologies, Sandia National Laboratories, Albuquerque, New Mexico 87175, USA

ABSTRACT Deterministic control over the location and number of donors is crucial to donor spin quantum bits (qubits) in semiconductor based quantum computing. In this work, a focused ion beam is used to implant antimony donors close to quantum dots. Ion detectors are integrated next to the quantum dots to sense the implants. The numbers of donors implanted can be counted to a precision of a single ion. In low-temperature transport measurements, regular coulomb blockade is observed from the quantum dots. Charge offsets indicative of donor ionization are also observed in devices with counted donor implants.

The spins of donor electrons in Si are promising candidates for quantum computing[1] because of long coherence times[2–4] and compatibility with existing fabrication technology. Single spin readout for donors in Si has been successfully demonstrated for P and Sb donors.[5,6] Entangled two-qubit operations, in addition to such single qubit operations, are required to form a universal set of quantum logic gates for circuit-based quantum computing.[7] However, many proposals for two-donor logic gates require stringent conditions on donor placement to be met as well as deterministic control over the number of donors.[8,9]

The first of these requirements, control of donor placement, can be achieved using two techniques – one is ion beam implantation and the second is hydrogen lithography using scanning tunneling microscopy (STM). The top-down technique has the advantage of being fast and compatible with existing fabrication processes at the cost of some uncertainty in ion placement. This uncertainty can be minimized by using a focused ion beam and by judicious choices of substrate, ion species and implantation energy.

The second requirement, control of the number of ions implanted, is especially difficult since ion implantation is a stochastic process. In previous experiments,[6,10] several devices were masked by poly methyl-methacrylate (PMMA) to limit implantation to certain areas. All devices were then exposed to an ion beam simultaneously. Standard implantation techniques were used to determine the total number of implanted ions which was then used to determine the 'average' number of ions implanted per device. This technique of determining the number of ions implanted, while excellent for a large number of ions, has inherent uncertainty levels which make it unfeasible when the desired number of ions implanted is reduced to 1 or 2. The use of a focused ion beam to target individual devices along with diode detectors to count the number of ions implanted is essential in this 'few donor' regime. Such diode detectors, with the ability to detect a single ion, have been demonstrated in proof of principle experiments[11–13] and alternatively to count dopants in conduction paths of a silicon nanostructure.[14,15] However, integrating these detection schemes into devices capable of electron spin readout and coherent control is a challenge. In this paper we present devices in which such integration has been achieved for the first time resulting in diode detectors for ion detection integrated adjacent to nanostructures for Si MOS-based single electron transistor (SET) formation. The device is built using current CMOS fabrication processes and is therefore integrable into current process flows. We successfully demonstrate detection of single ion impact events with these detectors and use this platform to synthesize devices with a counted number of implanted Sb donors. We observe electrostatically defined SET formation with regular coulomb blockade and charge offsets indicating the ionization of donors. This is the first demonstration of charge offsets in a counted donor device. The realization of a counted implant donor qubit platform opens an immediate path to fabricate two-donor devices which has been a goal of the donor qubit community for over a decade.

A scanning electron micrograph (SEM) of the device with the diode detector integrated to the left of it is shown in fig. 1a. A cross sectional schematic of the detector is shown in fig. 1b. The 'active' detection region of the diode detector is located between the P and the N electrodes. An electric field, intrinsic or applied, exists in the active region of the detector. When an ion strikes in this region, electron-hole pairs are generated and drift to the respective polarity terminals of the diode under the electric field. The 'construction zone', which is the area where the SET is patterned and the counted ion implants are targeted (see Fig. 1a), is located in the active detection region of the diode detector. Details of the fabrication process are included in the supplemental materials.

We first address the challenge of accurately placing a donor in a specific location. For ion implantation, we use a focused ion beam with a direct write lithography platform (nanoImplanter, A&D FIB100nI) that enables targeted single ion implantation into a chosen device. The nanoImplanter is also equipped with a low-temperature stage driven by laser interferometry, a fast blanking chopping mechanism and *in situ* electrical probes. The ion beam is generated using a liquid metal alloy ion source (LMAIS). The LMAIS in the nanoImplanter can be changed to provide an ion beam of a variety of materials like P, Bi and Co amongst others. In this study the LMAIS is an AuSiSb eutectic which can be tuned to an ion beam of Au, Si or Sb and the detection is done at room temperature. The accelerating voltage in conjunction with a Wien filter is used to select a particular ion with a specific energy. The beam spot size of the nanoImplanter depends on the type and energy of ion being used. The beam has a Gaussian intensity distribution. For a 120 keV Sb++ beam a 30 nm diameter spot measured at full-width-half-maximum (FWHM) was obtained. The polysilicon gates are approximately 60 nm wide. By positioning the beam over a polysilicon gate, self-alignment from the polysilicon electrode can be used to reduce lateral uncertainty to ~ 15 nm with 95% confidence. Additionally, PMMA masking by electron-beam lithography can be used to reduce lateral uncertainty even further.

The depth at which an ion comes to rest in the substrate is determined by the collisions the ion experiences and is a stochastic quantity. The width of the distribution of ion depths is known as the depth straggle and is determined by the species and energy of the ion as well as the material of the substrate. The depth straggle determines the vertical uncertainty in donor placement. Straggle in depth increases with increasing ion energy and decreases with increasing ion mass. To minimize the depth straggle, we choose to use low-energy, heavy Sb++ ions rather than much lighter P or As dopants. The depth straggle for 120 keV Sb++ ions in the $SiO_2$/Si substrate is 18nm, determined by stopping and range of ions in matter (SRIM)[17] simulations. The 120 keV Sb++ ions are on an average 37 nm below the $SiO_2$/Si interface, also from SRIM simulations. In comparison, for implanting light P+ ions at the same average depths, a 45 keV P+ ion beam would have to be used, resulting in a straggle of 26 nm. Depending on the angle of incidence and crystal orientation, some ions may not experience collisions with atoms in the lattice and thus, end up much deeper in the substrate than statistical collision simulations suggest. For the samples discussed here, the amorphous silicon dioxide on the surface of the Si substrate makes these 'channeling' effects negligible. We have calibrated the SRIM implantation depth estimates by time of flight secondary ion mass spectroscopy (SIMS) measurements. The samples for these measurements are implanted with 100 keV Sb ions at a fluence of 2 – 4 x $10^{12}$ ions/$cm^2$ in a 400 x 400 µ$m^2$ area. The depth distribution determined from SIMS is found to be 43 nm from the surface instead of 56 nm predicted by SRIM. SRIM calculations are known to produce systematic errors in stopping power calculations for low energy, heavy ions,[18,19] so this discrepancy is not unexpected. In addition to the reduced depth straggle, Sb is an attractive candidate for deterministic implantation for spin qubit applications since Sb does not segregate to the $SiO_2$/Si interface upon annealing.[20] After implantation, the samples are annealed at 900 C for 5 minutes. The intrinsic diffusion length for the heavy Sb ion under these anneal conditions is expected to be ~ 1.3 nm.[21]

As discussed before, diode detectors are integrated into the device to enable counted donor implantation. The breakdown characteristics of the diode detector[16] as a function of applied reverse bias are shown in Fig. 2a. The diode detector can be operated in the low bias linear mode or in the high bias Geiger mode. For the devices described here, the detector is operated at zero bias in linear mode. In linear mode, for a given energy, the number of electron-hole pairs generated is proportional to the number of ions incident in the active region for a given set of samples, ion species and energy. The detector response is therefore proportional to the number of ions implanted.

To detect the Sb ions, the electrical probes of the nanoImplanter are landed on the detector contacts. The probes are connected to a charge sensitive preamplifier (AmpTek A250 CoolFET) which provides an analog pulse with amplitude proportional to the integrated charge induced in the electrodes. The signal from the amplifier is then captured by an oscilloscope. The ion beam induced current (IBIC) flowing through the diode detector can thus be observed and recorded for each implantation event as it occurs. A single ion generates a very small induced current signal and requires a very sensitive diode detector and low noise electronics for detection. In our setup, a single 120 keV Sb implant through 35 nm $SiO_2$ generates 3150 e-h pairs according to SRIM. This produces a change of 8 mV in the detection signal at the oscilloscope. In fig. 2b, the high detection sensitivity is demonstrated with an IBIC map of a test device. In the test sample, the field oxide, the $Si_3N_4$ and the polysilicon (see fig. 3b for layer structure) are completely etched away from the construction zone using a reactive ion etch with an inductively coupled plasma source. The etch damages 1-2 nm of the oxide on the surface. This damaged oxide is removed using a dilute HF acid dip. We also use a 95% $N_2$, 5% $H_2$ forming gas anneal at 400°C for 30 minutes, which is known to suppress interface trap densities.[22] After the etching, a rectangular area with ~35 nm of thermal $SiO_2$ covering the Si substrate is left. When a 120 keV Sb++ beam is directed onto this sample, the beam can reach the substrate only in the 'exposed' construction zone and is stopped by the 450 nm thick field oxide or other layers in the remaining areas. A 2D scan of the region is performed by stepping the beam to each point, unblanking it for a set pulse length and recording the resulting detector response. The map in fig. 2b is obtained using an average of 1 ion/pulse beam scanning across the construction zone in 250 nm steps. As expected, the ions incident in the construction zone reach the active region of the diode and generate a detector response as shown in the figure. The ions outside the construction zone do not reach the substrate in the active region and there is no detector response. The average number of ions per pulse is adjusted to <N> = 1. According to the Poisson distribution, we expect 36.8% of the pulses to have 0 ions, 36.8 % to have 1 ion and the remaining 26.4% to have 2, 3 or more ions. From the response outside the construction zone, the average and standard deviation of the detector response to zero ions is obtained. A detector response higher than this 'noise' level corresponds to one or more implanted ions. Based on this calibration, the detector response to the 18921 ion pulses in the area marked by the dashed rectangle in the construction zone is analyzed and plotted in fig. 2c. The detector measures zero ions 39.5% of the time and one or more 60.5% of the time. The reason for there being ~ 3% more zeroes than predicted by the Poisson statistics could be that some of the ions do not make it through the thin remaining oxide, fabrication induced residue in some areas of the construction zone, damage from the fabrication generating 'dead' spots in the active detection region or slight fluctuations in beam current. One of the advantages of counting the number of ions implanted is that these problems are bypassed. Since the probability of 2 or more ions is only 26.4% and the detector is measuring a non-zero response 60.5% of the time, the detector is in fact sensitive to a single ion incident in the active region. Using this ability of the detector to detect a single ion, a counted single ion implant into the device can be performed. The procedure to produce devices with a high probability of single ion implants is to reduce the average number of ions per pulse to a small fraction of 1 ion per pulse so that Poisson statistics gives 0 ions in most pulses and 1 in a few. The probability of 2 or more ions at this low average number of ions per pulse is vanishingly small. The sample would be exposed to a large number of pulses until an ion strike is observed, guaranteeing a single ion implant with a high degree of confidence.

To confirm that the detector continues to work when integrated into devices that undergo nanoscale fabrication of a SET, another sample is made in which polysilicon is etched to form nanoscale gates defining a SET. In this device, the 120 keV Sb++ beam can only pass through the areas where the polysilicon has been etched and 35 nm of thermal $SiO_2$ is left (see fig. 3b). In other places the 200 nm polysilicon stops the ions. A scan similar to the one described before to obtain fig. 2b is performed with a 250 nm step size at 1 ion per pulse and an IBIC map of the device is generated (fig. 2d). This demonstrates that the detector is sensitive to single ions even after being integrated with the SET.

With the completion of devices with detectors integrated next to nanostructures, we next fabricated counted donor devices with multiple implants to demonstrate SET functionality and obtain evidence of the implanted donors in transport measurements. For these devices, a post-implant counting procedure in which the magnitude of the detector response is calibrated to the number of ions incident in the active region of the detector is used. Fig. 3a shows the calibration of the detector response for 120 keV Sb++ implants. To obtain this calibration, the pulse length of the ion beam is varied so that an increasing number of ions are incident in the construction zone in the active region of the detector (fig. 1a). For a given average number of ions, 10 pulses of a fixed pulse length are incident on the construction zone. The response of the detector for each pulse is recorded and the average response is plotted in Fig. 3a. The linear fit of voltage response dependence on number of ions is also shown in the figure. To produce a counted donor device, we set the average number of ions per pulse by knowing the beam current and setting the pulse length. We aim this pulse of known duration at the location in the construction zone where the implant is desired and record the resulting detector response. A calibration curve like the one shown in fig. 3a is used to determine the number of ions implanted into a device. Using this method of counting, a number of devices are implanted with a target number of 400, 100, 50, 25, 10 and 5 ions. An activation anneal and a forming gas anneal is performed on the samples after the implant (details in supplemental materials). This activation anneal is expected to activate 100% of the Sb donors.[3] The electrical connections to the various gates are made and the samples are cooled in a pumped helium system with a base temperature of 1.4 K.

Fig. 3b shows the cross sectional view of the device patterned in the construction zone and fig. 3c shows a SEM image with a top view of the polysilicon gates used to electrostatically define the SET. The wire gate (WG), left plunger (LP), central plunger (CP), right plunger (RP) and left (IsgL) and right isolation gates (IsgR) shown in fig. 3c are patterned in the 200 nm polysilicon layer of fig. 3b. By applying a positive voltage to WG, a 2D electron gas (2DEG) is formed at the Si/SiO$_2$ interface under the wire defined by WG. A negative voltage applied to LP,RP, IsgL or IsgR then depletes electrons from either end of the wire forming the dot and the tunneling junctions defining a SET. CP or any of the other gates can be used to vary the chemical potential of the dot. A small bias applied to the source and drain contacts, marked by crosses in fig. 3c and seen as the n+ doped region in fig. 3b, causes a current ($I_{SD}$) to flow through the SET. $I_{SD}$ is sensitive to changes in gate voltages and donor occupancy and can therefore be used to probe the donor electron dynamics.[6,10] The SET conductance is experimentally determined by applying an AC source-drain bias $V_{sd}$ = 100 µV and measuring the output current $I_{SD}$ using standard lock-in techniques at 441 Hz.

The implanted donors are capacitively coupled to the SET. Therefore, a change in the ionization state of a nearby donor leads to a discontinuity, or offset, in the charge transport through the SET. Several groups have reported this effect for traditionally implanted samples; that is, these experiments did not include a counted number of donors implanted next to a functional SET.

Fig. 3d shows $I_{SD}$ measured at 2 K as a function of WG and IsgL in a device with 35 nm oxide implanted with Sb++ ions at 120 keV in the implant window shown in fig. 3c. The device is found to have 27 ions implanted using the calibration curve shown in fig. 3a. A regular coulomb blockade pattern corresponding to the formation of an electrostatically defined quantum dot is visible. From the Coulomb blockade pattern shown here along with other stability diagrams, capacitances to various gates can be calculated. The capacitance of the SET to WG is found to be 9.2 aF, to CP 1.23 aF and to LP and RP ~ 1.54 aF and 0.41 aF respectively. These capacitance values are in agreement with those calculated for this device using a QCAD simulation[23] and are typical for other devices measured. A charge offset from an implanted donor is also visible in the figure. In 6 devices measured with similar implants, 10-50% of the counted implants are visible as charge offsets. In unimplanted devices, no charge offsets have been seen. Therefore we can form a lithographic SET in devices that have been integrated with a diode detector and detect counted ion implants as charge offsets in transport measurements.

In conclusion, we have developed devices with SET integrated next to diode detectors. We have demonstrated the ability to detect single ions impinging in the active region using this detector. We have fabricated a number of devices with a counted number of Sb donors at implant energies of 120 keV. Transport measurements performed on these devices demonstrate that an electrostatically defined dot can be formed and charge offsets from the implanted donors detected. The ability to implant a counted number of donors, in conjunction with the nanoImplanter gives us a direct path forward to synthesizing two donor devices. It must be stressed here that current technology, based on the statistical random distribution of ions and a large beam cannot control the number or the location of the ions while the technique we have developed using a focused ion beam and counting single ion implants can control both the number and the location. This control is necessary for future donor based spin qubit devices.

AUTHOR INFORMATION

**Corresponding Author**

*msingh@sandia.gov

**Author Contributions**

The manuscript was written through contributions of all authors. All authors have given approval to the final version of the manuscript.

ACKNOWLEDGMENT

This work was performed, in part, at the Center for Integrated Nanotechnologies, a U.S. DOE Office of Basic Energy Sciences user facility. Sandia National Laboratories is a multi-program laboratory operated by Sandia Corporation, a Lockheed-Martin Company, for the U. S. Department of Energy under Contract No. DE-AC04-94AL85000.

FIGURES

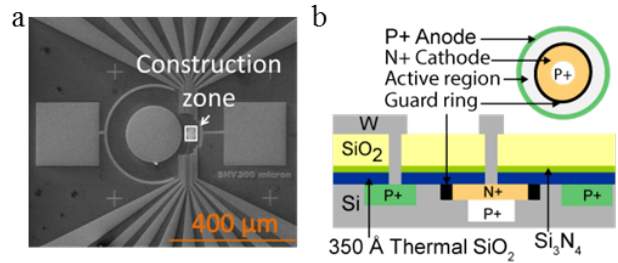

**Figure 1**. (a) SEM of a device including the construction zone, where the gates (light colored fan out pattern) to electrostatically define a dot are patterned and the diode detector used to count the number of ions implanted. The concentric circles are the contacts of the diode detector. (b) Schematic of the diode detector (top right) and the cross sectional view showing the various dopings and the 'active region' between cathode and anode of the diode where the construction zone is located. The single electron transistor for the transport measurements is fabricated in the construction zone in the active region. An SEM image of the device and its construction is shown in figure 3.

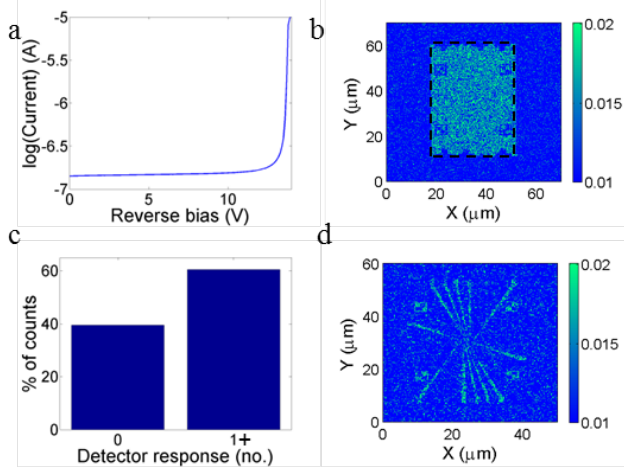

**Figure 2**. (a) Breakdown of a diode detector at 270 K (b) Test structure: Ion beam induced current map using a 120 keV Sb+ beam with 1 ion per pulse. The construction zone, which is the only area in which the stack is thin enough for the beam to reach the substrate, is also the only area that produces a response in the detector. The color represents the voltage pulse at the output of the charge sensitive amplifier. (c) Histogram showing the percentage of detector responses in the area within the dashed rectangle in 2(b) corresponding to zero ions incident (39.5%) or one or more ions incident (60.5%) out of a total of 18921 points. (d) Nanostructure: Ion beam induced current map showing the device in the construction zone generated using a 1 ion/pulse 120 keV Sb+ beam. The regions with the polysilicon gates stop the beam and appear blue in the figure. The regions where the polysilicon has been etched allow the beam to pass through and the detector responds corresponding to the green color in the figure.

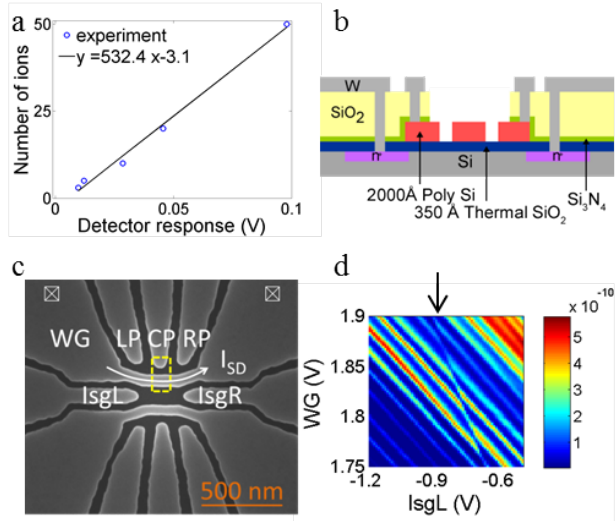

**Figure 3**. (a) Voltage response of diode detector to different number of (determined by pulse time) 120 keV Sb+ ions. Each point is an average of 10 data points. The line is the best linear fit. (b) Cross sectional view of single electron transistor shown in (c) including various dopings. The device is patterned in the construction zone in the active region shown in figure 1. (c) SEM showing two sets of polysilicon gates in the construction zone that can be used to form dots. The upper set are labelled, the source and drain denoted by ⊠ and the current path through the dot is shown. The dashed yellow window shows the location where the ion-implantation is done. (d) Transport at 2K through a device implanted with a counted 27, 120 keV Sb+ ions. Regular coulomb blockade and a charge offset can be seen.